\documentclass[aps,prb,reprint,showkeys]{revtex4-2}
\usepackage{graphicx}
\begin{document}

\title{First-principles calculations of vibrational spectra of CdSe/CdS superlattices}


\author{A. I. Lebedev}
\email[]{swan@scon155.phys.msu.ru}
%\thanks{}
\affiliation{Physics Department, Moscow State University, 119991 Moscow, Russia}

\date{\today}

\begin{abstract}
The vibrational spectra of CdSe/CdS superlattices (SLs) with different layer
thicknesses are calculated from first principles within the density functional
theory. It is shown that, along with folded acoustic and confined optical modes,
a number of confined acoustic modes appear in SLs. In structures with a minimum
thickness of one of the layers, microscopic interface modes similar to local
and gap modes in crystals appear. An analysis of projections of the eigenvectors
of vibrational modes in SLs onto the orthonormal basis of normal modes in binary
compounds enables to establish the details of formation of these vibrational modes
and, in particular, to determine the degree of intermixing of acoustic and optical
modes. A comparison of the frequencies of vibrational modes in CdSe/CdS SLs and
CdSe/CdS nanoplatelets enables to separate the influence of size quantization and
surface relaxation on the vibrational frequencies in the nanoplatelets.

\texttt{DOI: 10.21883/PSS.2022.14.54328.156 (Physics of the Solid State 64, 2270 (2022))}
\end{abstract}


\maketitle

\section{Introduction}

Vibrational spectroscopy techniques---Raman scattering and infrared (IR)
absorption---are powerful tools for analyzing the properties of various materials.
These methods have found a wide application in studies of low-dimensional
structures~\cite{PhononsSemiconductorNanostructures,YuCardona_engl,LightScatteringSolids.5.Chap3}.
They provide information on the real structure of the samples under study:
their composition, size, mechanical strains present in them, the state of the
interfaces, and surface relaxation of atoms. However, the interpretation of the
obtained results is often insufficiently substantiated. The aim of this work is
to analyze the vibrational spectra of CdSe/CdS superlattices (SLs) in order to
help in this interpretation. In the course of numerical modeling of the vibrational
spectra of SLs, we will encounter a number of previously little discussed vibrational
modes such as confined acoustic modes, gap modes, and local modes. In contrast to
a large number of previous works, in which simplified models were usually used
when calculating the lattice dynamics, in this work, we use the density functional
theory approach, in which the electrical and mechanical boundary conditions upon
relaxation of the structure (in our case, the lattice parameters of materials
differ by 4\%) are taken into account and satisfied automatically.

The vibrational spectra of superlattices are subjects of studies
for more than 40~years~\cite{ApplPhysLett.31.117,PhysRevB.17.3181,PhysRevLett.45.298,
ApplPhysLett.45.1138,PhysRevLett.54.2111,PhysRevLett.54.2115,PhysRevB.31.2080,
IEEEJQE.22.1760,JLumin.44.285,SuperlattMicrostruct.5.27,SuperlattMicrostruct.7.183,
JETP.71.603,PhysRevB.43.14754,PhysRevB.45.4280,PhysRevB.46.2375,PhysRevLett.72.1565,
PhysRevLett.73.740,SemicondSciTechnol.14.660,Nanotechnology.13.55,PhysRevB.69.132302,
Semiconductors.42.1208,PhysSolidState.54.1026,Nanomaterials.11.286}. Already in
the first papers, the main distinctive features of vibrational spectra of SLs have
been established:
the appearance in them of folded acoustic modes~\cite{ApplPhysLett.31.117,PhysRevB.17.3181,
PhysRevLett.45.298,PhysRevB.31.2080,IEEEJQE.22.1760,JLumin.44.285,SuperlattMicrostruct.5.27,
SuperlattMicrostruct.7.183} and confined optical modes~\cite{PhysRevLett.54.2111,
PhysRevB.31.2080,IEEEJQE.22.1760,JLumin.44.285,SuperlattMicrostruct.5.27,
SuperlattMicrostruct.7.183}.

The folded modes are longitudinal acoustic (LA) and transverse acoustic (TA)
vibrations that propagate in both materials, experiencing weak reflections at
the boundaries of two materials differing in their acoustic properties. The
frequency of such vibrations in two materials of the superlattice is the same,
but the wave vectors are different. The period of the superlattice determines new
periodic boundary conditions for the emergence of standing waves, and the fact
that the Brillouin minizone of the SL is several times smaller than the Brillouin
zone of bulk materials results in a folding of this zone in the growth direction so
that a number of points from the bulk of the Brillouin zones of the raw materials
are projected to the $\Gamma$~point of the folded zone. Thus, a whole set of new
vibrational modes appear in the SL at the center of the Brillouin zone. The
appearance of discontinuities in the energy spectrum of the modes (stop bands)
observed at the center and at the boundary of the Brillouin minizone is associated
with the difference in specific acoustic impedances of the used materials.

In the region of optical vibrations, confined modes, in which longitudinal optical
(LO) and transverse optical (TO) vibrations are localized in one of two materials
of the superlattice and are rapidly evanescent in another one, are observed. For
such modes to appear, it is necessary that vibrations with a given frequency can
propagate in one of the materials and cannot propagate in the other one. An
indication of such vibrations is the absence of mode dispersion and the strong
dependence of their frequency on the thickness of the layer of the first material.

Finally, under certain conditions, interface modes can arise in superlattices---modes
localized at the interface between two materials; these modes are evanescent in both
materials. Highly localized interface modes (they are also called \emph{microscopic})
were first discovered in InAs/GaSb~\cite{PhysRevB.33.8889} and Ge/Si~\cite{JPhysique.48.C5-569}
superlattices. Like the confined modes, these modes are characterized by the
absence of dispersion along the growth direction axis, but, in contrast to the
confined modes, their frequencies remain unchanged when changing the layer
thickness. We note that the appearance of interface modes depends on the polarization
of the vibrations: it is determined by how their frequencies relate to the frequencies
in the continua formed by longitudinal and transverse modes in bulk materials.

As follows from the experiment~\cite{PhysRevLett.54.2115,IEEEJQE.22.1760,JLumin.44.285,
SuperlattMicrostruct.5.27,SuperlattMicrostruct.7.183}, one more type of modes can
arise in superlattices. They are also referred to as interface modes, although they
are not strongly localized. These vibrational modes describe joint \emph{macroscopic}
vibrations of polar optical phonons in both materials of SL, in which the electric
fields generated by them are coupled by electrostatic boundary conditions. In the
limit of long waves, the frequencies of these modes satisfy the conditions
$d_1 \epsilon_1(\omega) + d_2 \epsilon_2(\omega) = 0$ or
$d_1 \epsilon_2(\omega) + d_2 \epsilon_1(\omega) = 0$, where $d_i$ is the thickness
of the $i$th layer in the SL, and $\epsilon_i$ is its complex dielectric
constant~\cite{ApplPhysLett.36.43,PhysRevLett.54.2115}. Since one of the dielectric
constants must be negative to fulfill these conditions, these frequencies fall into
the Reststrahlen band of one of the materials, that is, they lie
between the frequencies of TO and LO phonons in this bulk material. These modes can be
easily distinguished since their frequencies depend on the ratio of the layer thicknesses.
To observe these modes, studies of Raman spectra are usually carried out \emph{under resonance
conditions}~\cite{PhysRevLett.54.2115}. In contrast to the microscopic interface modes,
in which atomic vibrations are localized at the interfaces, a much larger number of
atoms usually participate in the vibrations of macroscopic interface modes (in the
long-wave limit, all atoms in both layers~\cite{PhysRevB.29.1695,PhysRevB.31.2080}).

The above results obtained for superlattices are also useful for understanding the
vibrational spectra of other quasi-two-dimensional
structures---nanoplatelets~\cite{PhysRevB.96.184306,JPhysChemC.123.11926,JPhysChemC.125.6758}.

\section{Calculation technique}

First-principles calculations were performed within the density functional theory
in the plane-wave basis and the local density approximation (LDA) using the
\texttt{ABINIT} software package~\cite{abinit3}. Optimized pseudopotentials for Cd, S,
and Se atoms were constructed according to the RRKJ scheme~\cite{PhysRevB.41.1227}
using the \texttt{opium} program. The maximum energy of plane waves in the calculations
was 30~Ha (816~eV). Integration over the Brillouin zone was carried out using the
8$\times$8$\times$4 or 8$\times$8$\times$2 Monkhorst--Pack meshs. The lattice
parameters and equilibrium positions of atoms in superlattices oriented in the [001]
direction and containing up to twelve monolayers of semiconductor were obtained from
the condition that the Hellmann--Feynman forces become less than $5 \cdot 10^{-6}$~Ha/Bohr
(0.25 meV/{\AA}) while the accuracy of calculating the total energy is better than
10$^{-10}$~Ha. The phonon spectra of the obtained equilibrium structures were calculated
using the density-functional perturbation theory analogously to~\cite{PhysSolidState.51.362}.

In this work, when determining the nature of vibrational modes, we will analyze
the dispersion of these modes along the $\Lambda$~axis of the Brillouin zone since
in real space it corresponds to the $z$~direction in which strong perturbations are
created in the superlattice structure. Fortunately, a high symmetry of the little group
of the $\Lambda$ wave vector retains the division of vibrations into the longitudinal
and transverse ones. We will not try to establish whether these modes belong to
macroscopic interface modes, since for this it is necessary to analyze their properties
at nonzero transverse component of the phonon wave vector.

\section{Calculation results and their discussion}

\subsection{Phonon spectra of superlattices}

\begin{figure}
\includegraphics{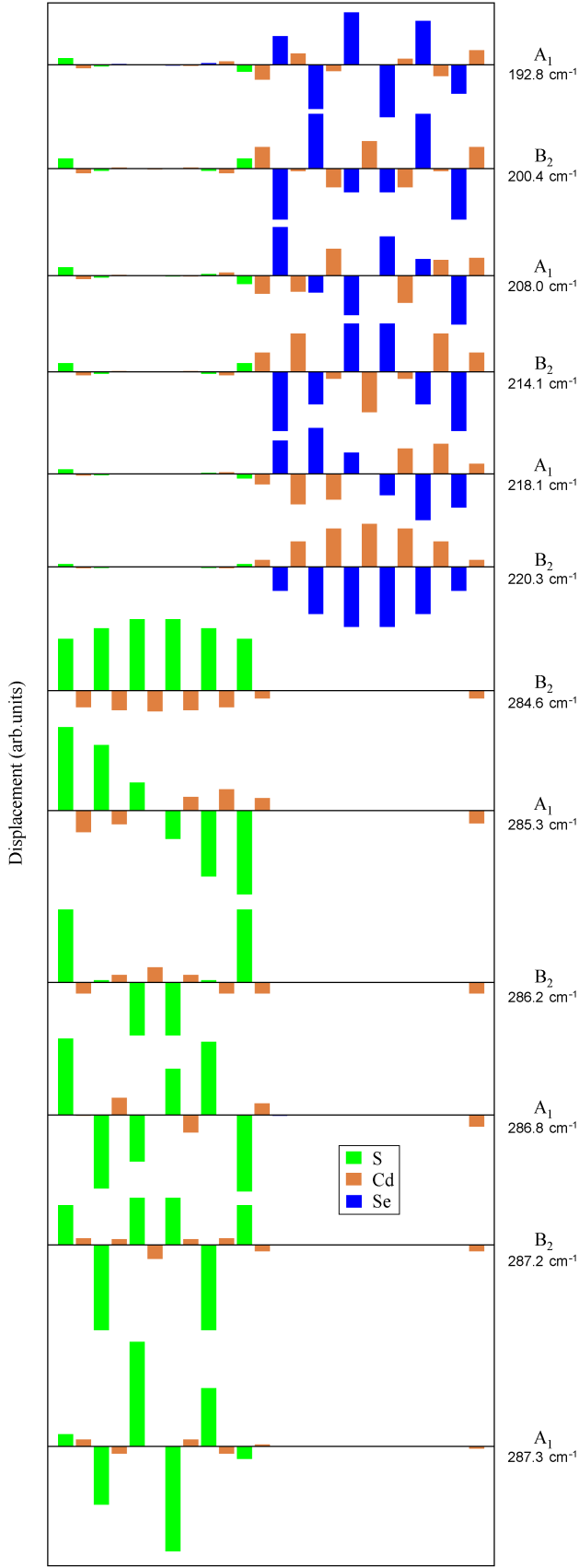}
\caption{\label{fig1}Eigenvectors and frequencies of longitudinal optical modes
with the $A_1$ and $B_2$~symmetry at ${\bf q} \to 0$ in the (CdSe)$_6$(CdS)$_6$
superlattice.}
\end{figure}

\begin{figure}
\includegraphics{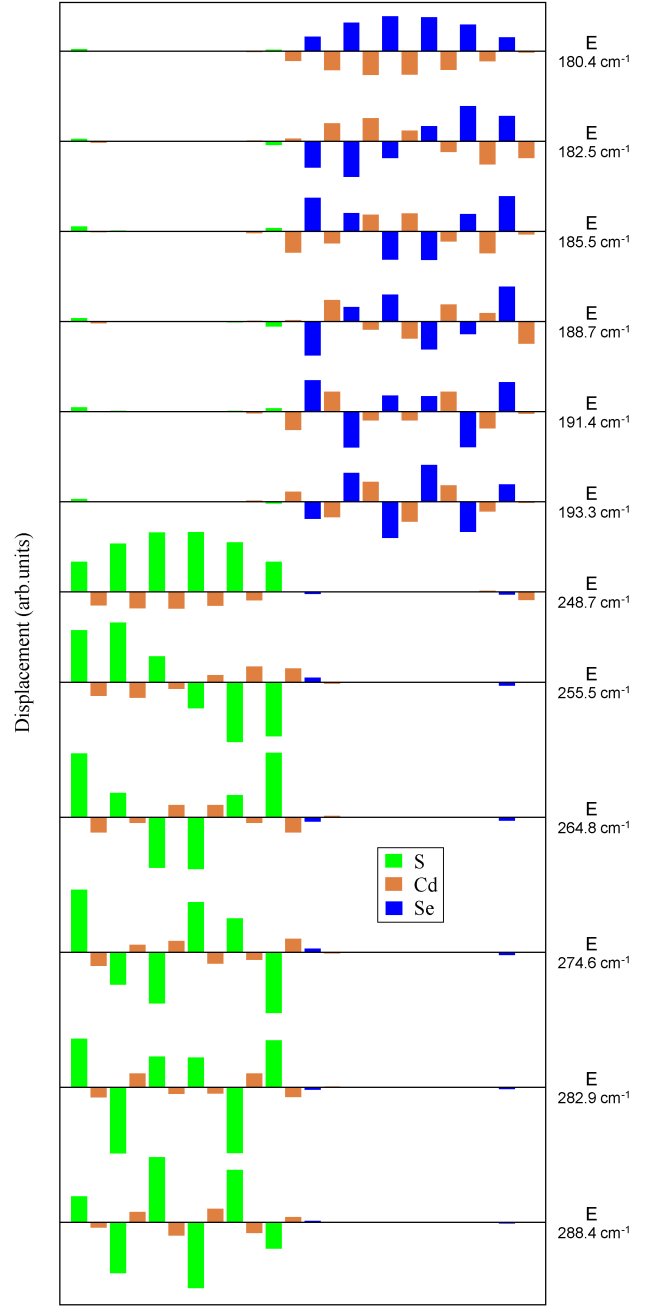}
\caption{\label{fig2}Eigenvectors and frequencies of transverse optical modes of
the $E$~symmetry at ${\bf q} = 0$ in the (CdSe)$_6$(CdS)$_6$ superlattice.}
\end{figure}

The symmetry of all superlattices studied in this work is described by the $P{\bar 4}m2$
space group, and the phonon modes at the $\Gamma$~point can have the $A_1$, $B_2$, and
$E$ symmetry.  The eigenvectors of all optical modes for the (CdSe)$_6$(CdS)$_6$
SL as well as their frequencies and symmetries are shown in Figs.~\ref{fig1} and
\ref{fig2}. As follows from the figures, all obtained optical modes are confined:
the vibrations are localized in one material of the SL and rapidly decay in the second
material.

\begin{figure*}
\includegraphics{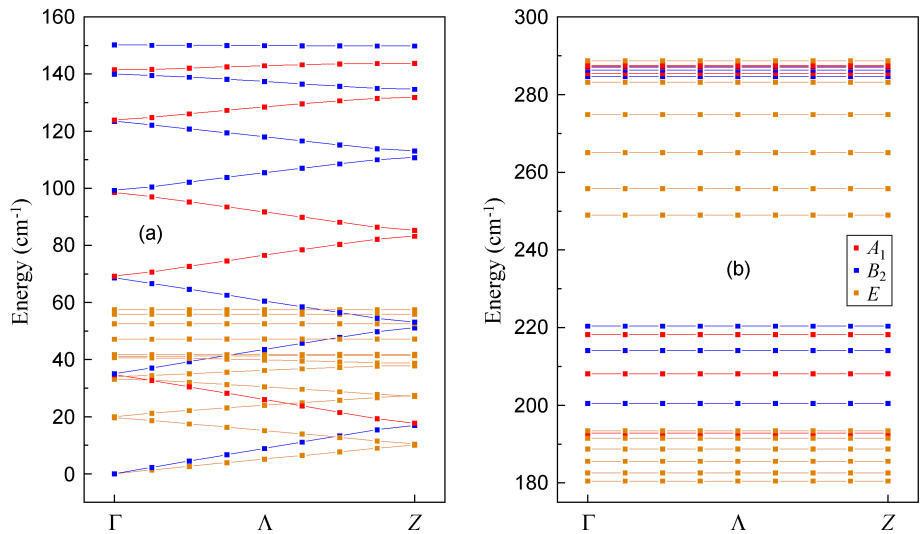}
\caption{\label{fig3}Dispersion curves of (a) acoustic and (b) optical phonons
in the (CdSe)$_6$(CdS)$_6$ superlattice.}
\end{figure*}

An analysis of the dispersion curves for optical modes (Fig.~\ref{fig3}) shows that in
the $\Gamma$--$Z$ direction the frequencies of these modes are practically independent
of the wave vector (changes in the mode frequencies do not exceed 0.01~cm$^{-1}$).
This behavior is consistent with the existing concept of confined modes: they do
not exhibit dispersion, but their frequencies depend on the thickness of the SL
layers. The latter was demonstrated on (CdSe)$_n$(CdS)$_n$ SLs with individual layer
thicknesses of $n = {}$1--4.

When discussing the behavior of dispersion curves in the Brillouin zone of a
tetragonal structure, the following should be borne in mind. When moving from
the $\Gamma$~point to the $\Lambda$~axis, the compatibility relations of
irreducible representations describing the symmetry of normal vibrations are
$A_1 \to \Lambda_1$, $B_2 \to \Lambda_1$, $E \to \Lambda_3 + \Lambda_4$
($\Lambda_3$ and $\Lambda_4$ are conjugate representations degenerate in
frequency)~\cite{bilbao}. Thus, on the $\Lambda$~axis, only longitudinal $A_1$
and $B_2$~modes can mix with each other, although the mixing of acoustic and optical
modes is also possible. When moving from the $\Gamma$ point along the $\Delta$ and
$\Sigma$~axes of the Brillouin zone, the situation is different: the compatibility
relations $A_1 \to \Delta_1$, $B_2 \to \Delta_1$, $E \to \Delta_1 + \Delta_2$ and
$A_1 \to \Sigma_1$, $B_2 \to \Sigma_2$, $E \to \Sigma_1 + \Sigma_2$ allow mixing of
longitudinal and transverse vibrations. In this case, significant dispersion appears
on the dispersion curves, and the $E$~modes are strongly split. The component of the
$E$~mode, which transforms according to the $\Delta_2$ representation, does not mix
with the $A_1$ and $B_2$~modes polarized in the $z$~direction, and therefore its
eigenvector describes only the displacements ${\bf u} \perp {\bf q}$ in the $xy$~plane.
The $\Delta_1$~mode as well as both $\Sigma_1$ and $\Sigma_2$ modes experience
intermixing and exhibit complex displacement patterns in all three directions.

An analysis of the dispersion curves along the $\Lambda$~axis in the region of
acoustic vibrations finds folded longitudinal and transverse acoustic modes
(Fig.~\ref{fig3}) and the appearance, in the region of 45--60~cm$^{-1}$, of
several dispersionless TA modes, which will be discussed in Sec.~\ref{sec3.4}.

Microscopic interface modes, whose eigenvectors are localized at the interface
and which decay when moving into the interior of both materials, are not observed
in CdSe/CdS SLs. Apparently, in superlattices whose materials have one common atom,
this cannot be obtained in principle because the frequencies of potential interface
modes in these SLs fall into the continuum of optical vibrations of at least one of
the materials of the SL. However, a kind of a microscopic interface mode can be
created artificially if we consider the properties of SLs with extremely thin layers
of one of the materials.

\subsection{Local and gap interface modes}
\label{sec3.2}

The microscopic interface mode in a superlattice can be obtained from the confined
mode when the minimal thickness (one monolayer) of one of materials is used. If
the frequency of this mode is outside the continuum of optical modes of the matrix,
then such vibrations will not be able to propagate in the matrix and will be localized.

\begin{figure*}
\includegraphics{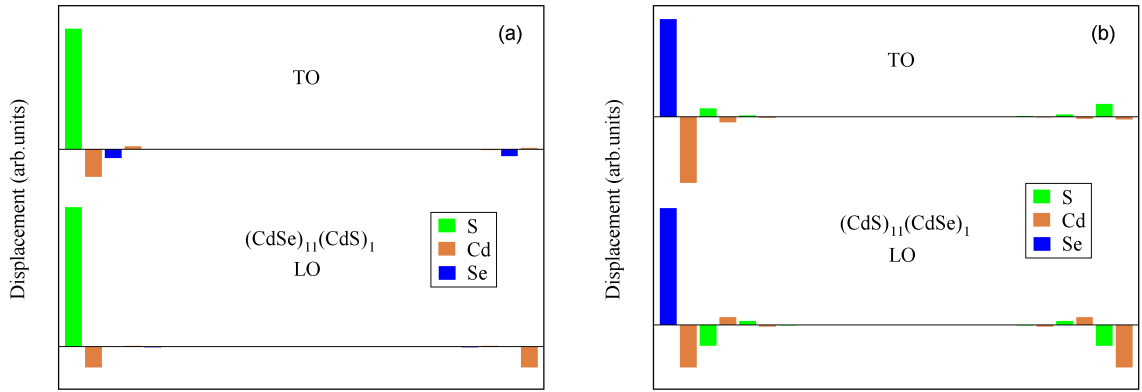}
\caption{\label{fig4}Eigenvectors of the (a) local and (b) gap LO and TO modes at
${\bf q} \to 0$ in (CdSe)$_{11}$(CdS)$_1$ and (CdSe)$_1$(CdS)$_{11}$ superlattices.}
\end{figure*}

We performed a search for such modes in the (CdSe)$_{11}$(CdS)$_1$ and (CdSe)$_1$(CdS)$_{11}$
SLs, which contain one CdS monolayer in the CdSe matrix and one CdSe monolayer in the
CdS matrix, respectively. As expected, localized optical modes of S and Se vibrations
arised in the structures. In both cases, the incorporation of a monolayer into
a matrix gives rise to three split-off modes whose symmetry ($B_2 + E$) coincides
with the symmetry of optical phonons in bulk materials ($\Gamma_{15} \to B_2 + E$).
In the CdSe matrix, the vibration frequencies of the S layer are, respectively,
265.2 and 274.6~cm$^{-1}$; they lie above the upper limit of the optical modes
continuum of the matrix and are local modes (we use here a terminology used for
classification of localized vibrations of impurities in crystals). In the CdS matrix,
the vibrational frequencies of the Se layer are 187.3 and 217.0~cm$^{-1}$, respectively;
they lie in the gap between the optical and acoustic modes continua and are the gap
modes. The eigenvectors of the local and gap modes are shown in Fig.~\ref{fig4}. The
admixture of acoustic vibrations to the transverse local \emph{optical} mode reaches
20\% at the $Z$~point, and that to the transverse gap mode reaches 45\% at the
$Z$~point. In the local and gap longitudinal modes, the admixture of acoustic
vibrations does not exceed 1\%.

As expected for the local modes, the dispersion of both the $E$ and $B_2$~modes along
the $\Lambda$~axis in the (CdSe)$_{11}$(CdS)$_1$ SL is absent. When the thickness
of the CdSe layer is decreased from 11 to 7~monolayers, the mode frequencies exhibit
a slight change (by 0.16 and 0.41~cm$^{-1}$, respectively), which is entirely due to
a small (by 0.2\%) change in the in-plane lattice parameter of the SL. The gap
modes are also dispersionless, and their frequencies change by no more than
0.002~cm$^{-1}$ when decreasing the number of CdS monolayers from 11 to 7 at a
fixed lattice parameter.

While the local modes are well known, the gap modes have been studied in less
detail. The possibility of their appearance turns out to be more problematic since
for this the frequencies of these modes must fall into the gap between the acoustic
and optical modes continua of both materials of the SL. For transverse vibrations
in CdSe/CdS SLs, the upper limit of the acoustic modes continuum lies in the
region of 60 cm$^{-1}$, and the gap under discussion is rather large. For
longitudinal vibrations, the possibility of the appearance of a gap mode is
determined by the maximum frequency of the LA mode at the $X$~point in CdS, which is
152.2~cm$^{-1}$. We note, however, that because of the orthogonality of longitudinal
and transverse modes everywhere on the $\Lambda$~axis, the possibility of observing
transverse gap modes turns out to be wider. For example, the \emph{transverse} In--Sb
gap mode observed in the GaSb/InAs SLs in Ref.~\cite{PhysRevB.33.8889} remained
dispersionless (i.e., localized) despite the fact that it was superimposed on
the \emph{longitudinal} optical modes continuum for one of materials of the SL
(InAs).

Calculations show that the contributions of the discussed localized modes to
Raman and IR spectra are large enough to be observed experimentally.

\subsection{Dependence of TO-mode frequencies on the superlattice period}

\begin{figure}
\includegraphics{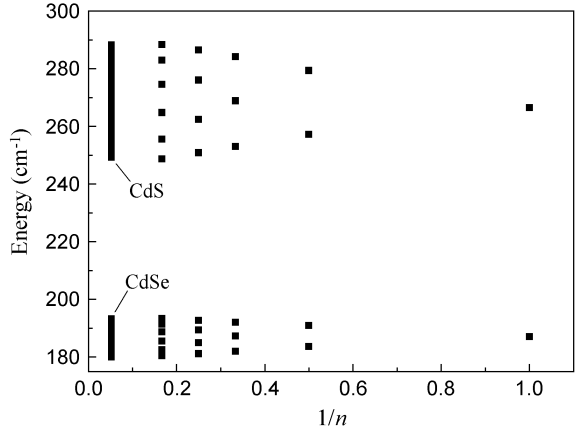}
\caption{\label{fig5}Frequencies of TO modes at ${\bf q} = 0$ in (CdSe)$_n$(CdS)$_n$
superlattices. The wide vertical lines show the ranges of TO mode frequencies in
CdSe and CdS binary compounds.}
\end{figure}

In Ref.~\cite{JPhysChemC.125.6758}, when studying the vibrational spectra of CdSe/CdS
nanoplatelets, we discovered a mode associated with TO vibrations in the CdS layer,
whose frequency rapidly decreased with increasing thickness of this layer. This mode
gave the strongest contribution to the IR spectra from the CdS layers. The observed
effect was explained by surface relaxation of the structure, which resulted in a
noticeable shortening of the Cd--S interatomic distances near the surface and a
corresponding increase in the vibrational frequency. The appearance of an $E$~mode
with similar properties in CdSe/CdS superlattices (lowest points in the upper block
of frequencies in Fig.~\ref{fig5}) and an analysis of the relaxations of the Cd--S
distances (as compared to those in bulk CdS) when changing the period of
(CdSe)$_n$(CdS)$_n$ SLs showed that the maximum change in these distances in SLs
is an order of magnitude smaller than in nanoplatelets, and, moreover, the changes
in SLs have an opposite sign (Table~\ref{table1}). This may mean that the mechanism
proposed in~\cite{JPhysChemC.125.6758} is not the only one.

\begin{table}
\caption{\label{table1}Frequencies of TO modes giving the strongest contribution
to the IR spectra from the CdS layers and the relaxation of the average Cd--S bond
length for (CdSe)$_4$/(CdS)$_n$ nanoplatelets and (CdSe)$_n$(CdS)$_n$ superlattices
with different number of monolayers $n$ in the CdS layer. All calculations were
performed using the same pseudopotentials.}
\begin{ruledtabular}
\begin{tabular}{ccccc}
$n$ & \multicolumn{2}{c}{Mode frequency (cm$^{-1}$)} & \multicolumn{2}{c}{Relaxation $R_{\rm Cd-S}$} \\
    & Nanoplatelet & Superlattice & Nanoplatelet & Superlattice \\
\hline
1   & 270.5 & 266.5 & $-$0.758\% & +0.522\% \\
2   & 259.8 & 257.2 & $-$0.248\% & +0.485\% \\
3   & 255.0 & 253.1 & $-$0.114\% & +0.478\% \\
4   & ---   & 250.9 & ---        & +0.473\% \\
6   & ---   & 248.8 & ---        & +0.470\% \\
\end{tabular}
\end{ruledtabular}
\end{table}

In order to understand the origin of the strong effect of the thickness of CdS
layer on the TO mode frequency in CdSe/CdS SLs, we analyzed the projections
of the eigenvectors of vibrational modes in SLs by expanding them in the
orthonormal basis of normal vibrations of LA, LO, TA, and TO phonons of bulk CdS
with the zinc-blende structure:
\begin{equation}
{\bf Q}^{\rm SL}_{\lambda} = \sum_{n=1}^{4} \sum_{q} C^{\lambda}_{nq} {\bf Q}^{\rm bulk}_{nq}.
\end{equation}
Here $\lambda$~is the number of the vibrational mode, and $q$~is the wave vector
of the normal mode. To make use of the orthogonality of normal modes, we should
work with eigenvectors of the dynamical matrix ${\bf Q}_{nq}$, which are obtained
by componentwise multiplication of the displacement vectors of the normal modes
${\bf u}_{nq}$ by square roots of the masses of corresponding atoms.

The basis of the normal modes in bulk CdS was constructed according to a scheme
similar to that used in Ref.~\cite{PhysRevB.96.184306}. First, the eigenvectors of
LA, LO, TA, and TO~phonons were calculated from first principles at 13~points of the
Brillouin zone for dimensionless wave vectors $0 \le q_z \le 1$ located between the
$\Gamma$ ($q_z = 0$) and $X$ ($q_z = 1$) points of the Brillouin zone. The ratio of
the displacements of Cd and S atoms for these phonons was
approximated by a fourth-order polynomial of the $\sqrt{\cos(\pi q_z/2)}$ function
for longitudinal modes and the $\cos^2(\pi q_z/2)$ function for transverse modes.
These polynomials were then used to construct normalized basis functions for arbitrary
value of $q_z$. For different $q_z$, the basis functions are orthogonal by construction.
Checking the basis functions for the same $q_z$ showed that the deviation from their
orthogonality does not exceed 0.003 for all $q_z$ values.

\begin{figure*}
\includegraphics{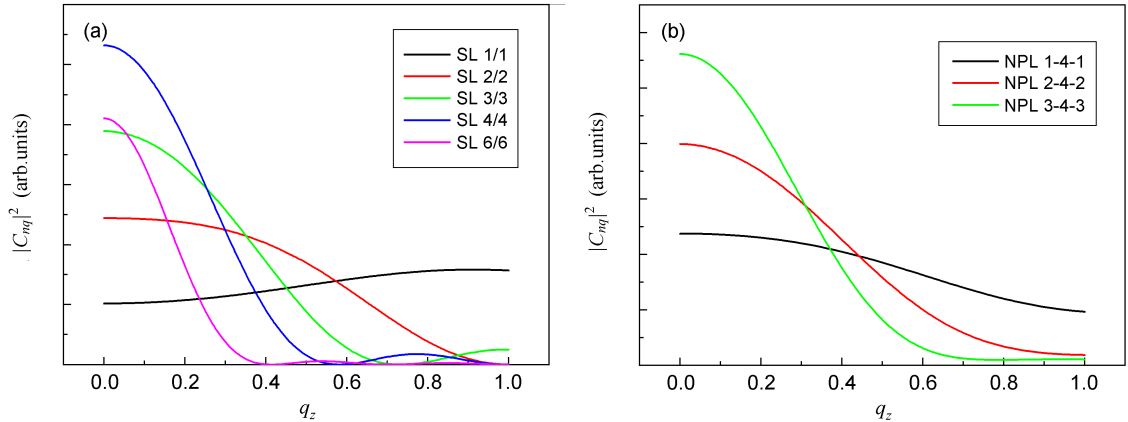}
\caption{\label{fig6}Projections of the eigenvectors of TO modes giving the largest
contribution to the IR spectra from the CdS layers onto the eigenvectors of TO phonons
in bulk CdS with wave vectors between the $\Gamma$ and $X$~points for (a)
(CdSe)$_n$(CdS)$_n$ superlattices and (b) (CdSe)$_4$/(CdS)$_n$ nanoplatelets.}
\end{figure*}

The squared moduli of the $C^{\lambda}_{nq}$ coefficients as a function of $q_z$
for TO modes in SLs are shown in Fig.~\ref{fig6}(a). The largest contribution
to the displacement patterns of these modes is made by TO phonons of CdS; the
contribution of TA phonons is about 100~times smaller. For the SL with the smallest
thickness of the CdS layer, TO phonons from the entire Brillouin zone of bulk CdS
contribute to this TO mode. Therefore, it is not surprising that the frequency of
this mode in the (CdSe)$_1$(CdS)$_1$ SL is close to the average frequency of TO
vibrations in bulk CdS. With an increase in the thickness of the CdS layer, phonons
from an increasingly narrow range of wave vectors near the $\Gamma$~point start
to dominate in the contributions. This, in accordance with the dispersion curve
for TO phonons in bulk CdS, results in a decrease in the frequency of the optical mode.

The projection analysis of the displacement patterns for TO modes in (CdSe)$_4$/(CdS)$_n$
nanoplatelets shows that the range of wave vectors contributing to these modes is
noticeably narrower than in (CdSe)$_n$(CdS)$_n$ superlattices with the same thickness
of the CdS layer (Fig.~\ref{fig6}). If the size quantization effect were the only one,
the mode frequency in nanoplatelets would be lower than in superlattices. The fact that
the opposite effect is actually observed (Table~\ref{table1}) means that in nanoplatelets
there exists one more contribution, namely the surface relaxation discussed
in~\cite{JPhysChemC.125.6758}.

For the local TO mode in the (CdSe)$_{11}$(CdS)$_1$ SL considered in Sec.~\ref{sec3.2},
the projection onto TO phonons in CdS (Fig.~\ref{fig6}(b)) is very similar to the
projection for the (CdSe)$_1$(CdS)$_1$ SL. However, in this case the contribution of
TA modes becomes quite noticeable (21\% in the vicinity of the $X$~point). For the
local LO mode in the (CdSe)$_{11}$(CdS)$_1$ SL, the dominant contribution to this
vibration comes from LO phonons of CdS from the vicinity of the $X$~point, but the
fact that the frequency of this mode in the SL (274.6~cm$^{-1}$) is much lower than
the frequency of LO~phonon at the $X$~point in CdS (300.6~cm$^{-1}$) may indicate a
noticeable contribution of CdSe to this mode. The observed frequency shift is
similar to the frequency shift for local vibrations of isolated impurities (for
longitudinal local vibrations of the S impurity in the CdSe matrix, the calculated
frequency is 280.3~cm$^{-1}$).

\subsection{Confined TA modes}
\label{sec3.4}

In principle, the possibility of the appearance of confined TA modes arises from
the difference in the spectra of acoustic vibrations in two materials of the superlattice.
In particular, such modes were observed in GaSb/InAs SLs~\cite{PhysRevB.33.8889},
where they were associated with the local vibrations of the Ga--As atomic pair located
at the interface. The interface character of this mode was supported by the
independence of its frequency from the thickness of layers in the superlattice.
Similar modes did not appear at the In--Sb interface, since their frequencies fell
into the continuum of acoustic modes of bulk materials.

\begin{figure}
\includegraphics{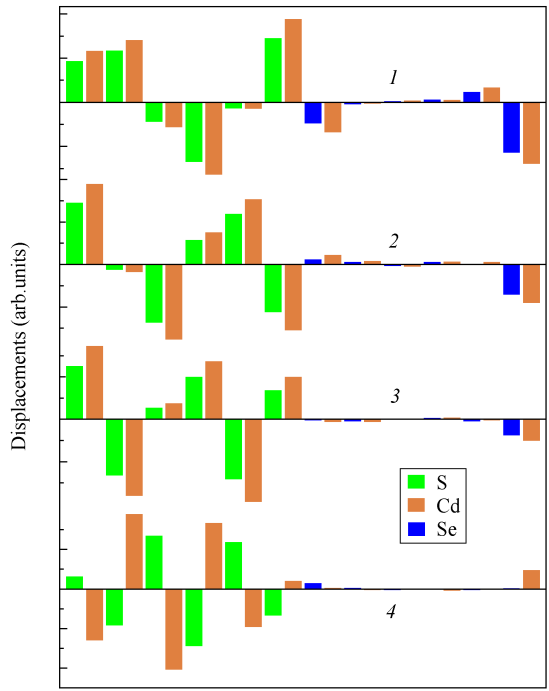}
\caption{\label{fig7}Eigenvectors of four confined TA modes with frequencies of (1) 47.9,
(2) 53.4, (3) 56.8, and (4) 58.3 cm$^{-1}$ (${\bf q} = 0$) in the (CdSe)$_6$(CdS)$_6$
superlattice.}
\end{figure}

An analysis of the acoustic modes in the phonon spectrum of the (CdSe)$_6$(CdS)$_6$ SL
finds in it, according to the vibration eigenvectors, four \emph{confined} TA vibrational
modes with frequencies of 47.9, 53.4, 56.8, and 58.3~cm$^{-1}$ (Fig.~\ref{fig7}). Similar
modes were found in superlattices (CdSe)$_{11}$(CdS)$_1$ (frequency 52.0~cm$^{-1}$),
(CdSe)$_{10}$(CdS)$_2$ (frequencies 46.8 and 56.6~cm$^{-1}$), and (CdSe)$_9$(CdS)$_3$
(frequencies 51.7 and 58.2~cm$^{-1}$). According to our estimates, the number of such
modes in long-period SLs is about 70\% of the number of CdS layers. Calculations of
the dispersion curves along the $\Lambda$~axis for these modes in (CdSe)$_6$(CdS)$_6$
and (CdSe)$_{11}$(CdS)$_1$ SLs show that the modes are dispersionless (the frequency
change does not exceed 0.01~cm$^{-1}$, Fig.~\ref{fig3}). Comparison of the frequencies
of these modes with those of TA phonons at the $X$~point in bulk crystals (45.0~cm$^{-1}$
in CdSe and 55.0~cm$^{-1}$ in CdS) shows that transverse vibrations with such frequencies
indeed cannot propagate in the CdSe layers and, therefore, are localized in the CdS layers.
It is interesting that the frequencies of several of these modes even exceed the frequency
of the upper limit of the acoustic continuum in bulk CdS. This may be due, first, to an
admixture of up to 17--24\% of optical vibrations to these acoustic modes, and second,
to an increase in the frequency of TA phonons in the CdS layer upon biaxial stretching
of the structure, which results from the addition of CdSe layers.

\begin{figure*}
\includegraphics{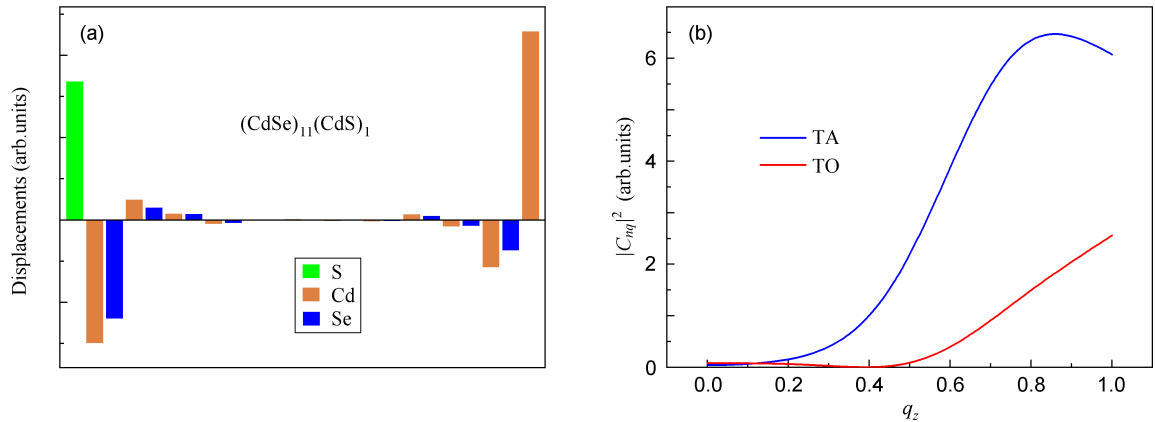}
\caption{\label{fig8}(a) Eigenvector of the interface TA mode in the (CdSe)$_{11}$(CdS)$_1$
superlattice. (b) Projections of the eigenvector of this mode onto the normal TA and
TO modes of bulk CdS.}
\end{figure*}

In SLs with a shorter period ((CdSe)$_4$(CdS)$_4$), it is more difficult to draw
conclusions about the nature of a mode from its eigenvector; however, the calculation
of the mode dispersion along the $\Lambda$~axis indicates that the mode with
an energy of 56.9~cm$^{-1}$ is dispersionless, and modes with frequencies of
53.1 and 45.9~cm$^{-1}$ have a small dispersion (0.03 and 0.13~cm$^{-1}$, respectively),
which indicates a weak interaction of vibrations in neighboring CdS layers in this SL
(recall that the decay rate of vibrations increases when moving away from the upper
limit of the acoustic modes continuum in the CdSe matrix, which is 45.0~cm$^{-1}$).
Since the vibrational frequencies of the discussed TA modes depend on the layer thickness,
these modes are confined but not interface modes. Estimates of the contribution of
the considered TA modes to the infrared and Raman spectra show that the experimental
observation of these modes can be problematic since their contribution to both types
of spectra is rather small.

In continuation of our discussion of localized vibrations in SLs with one extremely
thin layer, it is interesting to discuss whether the confined TA~mode with a frequency
of 52.0~cm$^{-1}$ arising in the (CdSe)$_{11}$(CdS)$_1$ SL (Fig.~\ref{fig8}) can be
interpreted as an \emph{interface} mode. In contrast to microscopic interface optical
modes, the peculiarity of acoustic vibrations is that their eigenvectors have a double
structure associated with the simultaneous occurrence of stretching and bending bond
deformations that accompany lattice vibrations. The authors of Ref.~\cite{PhysRevB.33.8889}
considered the TA mode with similar properties (resulting from the Ga--As pair vibrations
in the GaSb/InAs SLs) as an interface mode. As in that work, the mode we are discussing
is detached from the continuum of folded modes, is localized, and has no dispersion along
the $\Lambda$~axis. Very small (by 0.17~cm$^{-1}$) shift of the frequency of this mode
when the thickness of the CdSe layer in the superlattice is decreased from 11 to 7~monolayers
is entirely caused by a change in the in-plane lattice parameter of the SL and confirms
the conclusion about its interface character. According to the projection analysis
(Fig.~\ref{fig8}(b)), the largest contribution to the discussed TA mode is given by
TA phonons from a wide vicinity of the $X$~point of bulk CdS with a noticeable (up to
30\%) admixture of TO phonons (the same symmetry  of these vibrations allows their mixing).

\section{Conclusions}

In this work, the vibrational spectra of CdSe/CdS superlattices (SLs) are calculated
from first principles within the density functional theory. It is shown that along
with folded acoustic and confined optical modes, a whole set of confined acoustic
modes, whose number is $\sim$70\% of the number of layers of material with a higher
frequency of TA phonons, appears in the SLs. In structures with a minimum thickness of
one of the layers, the formation of microscopic interface modes such as local and gap
modes is possible. An analysis of the projections of the eigenvectors of vibrational
modes in SLs onto the orthonormal basis of normal modes in binary compounds finds a
fairly intense mixing of acoustic and optical vibrations even in modes traditionally
referred to as acoustic or optical.

%\appendix
%\section{}

%\begin{acknowledgments}
%\end{acknowledgments}

%\bibliography{all}
\providecommand{\BIBYu}{Yu}

\end{document}